# Examining Student Interactions with a Pedagogical AI-Assistant for Essay Writing and their Impact on Students' Writing Quality

Wicaksono Febriantoro[*]

University College London (UCL), wicaksono.febriantoro.21@ucl.ac.uk

Qi Zhou

University College London (UCL), qtnvqz3@ucl.ac.uk

Wannapon Suraworachet

University College London (UCL), wannapon.suraworachet.20@ucl.ac.uk

Sahan Bulathwela

University College London (UCL), m.bulathwela@ucl.ac.uk

Andrea Gauthier

University College London (UCL), andrea.gauthier@ucl.ac.uk

Eva Millan

Universidad de Málaga, emillan@uma.es

Mutlu Cukurova

University College London (UCL), m.cukurova@ucl.ac.uk

The dynamic nature of interactions between students and GenAI, as well as their relationship to writing quality, remains underexplored. While most research has examined how general-purpose GenAI can support writing, fewer studies have investigated how students interact with pedagogically designed systems across different phases of the writing process. To address this gap, we evaluated a GenAI-driven essay-writing assistant (EWA) designed to support higher education students in argumentative writing. Drawing on 1,282 interaction logs from 32 undergraduates during a two-hour writing session, Sequential Pattern Mining and K-Means clustering were used to identify behavioral patterns. Two clusters emerged: Cluster 1 emphasized outline planning and essay structure, while Cluster 2 focused on content development. A Mann-Whitney U test revealed a moderate effect size ($r = 0.36$) in the essay Organization dimension, with Cluster 1 showing higher scores. Qualitative analysis indicated that students with better performance actively wrote and shared essay sections with EWA for feedback, rather than interacted passively by asking questions. These findings suggest implications for teaching and system design. Teachers can encourage active engagement, while future EWAs may integrate automatic labeling and monitoring to prompt students to move from questioning to writing, enabling fuller benefits from GenAI-supported learning.

---

[*] Place the footnote text for the author (if applicable) here.

**ACM Reference Format:**

# 1 INTRODUCTION

Research has shown Generative AI (GenAI) potential in enhancing learning outcomes in various educational context, including academic writing, yet its effectiveness depends on learners' critical assessment of AI-generated responses [19, 31]. As off-the-shelf GenAI has been trained on large but unspecified datasets, it was not designed specifically for education. This raises concerns about inaccuracies [1] and overly generalized responses [10], suggesting the need for context-specific alignment. There is also a risk of students' over-reliance on GenAI tools which may undermine their creativity [3] and independent thinking [25] as well as reducing learner's ability to independently retain and transfer knowledge [32]. One strategy to mitigate these risks is to employ pedagogical prompts, equipping GenAI to act like an effective tutor that stimulates active engagement with learning materials [8, 11]. Pedagogical prompting refers to a prompting technique used to instruct GenAI to respond to end-users (students) in a pedagogically informed manner, imitating effective tutor characteristics such as those outlined in the INSPIRE model [17], which identifies Intelligent, Nurturant, Socratic, Progressive, Indirect, Reflective, and Encouraging traits. This approach reinforces students' critical thinking by offering guiding questions rather than direct answers [6]. This study further extends the application of the INSPIRE model [17] and effective tutoring behavior [6], which are currently implemented by human tutors in in-person and online settings respectively, by adapting these models for GenAI implementations.

This vision has fueled growing interest in the potential of GenAI in assisting academic writing tasks. By integrating academic writing theory [9] and formative feedback principles [18, 23], a GenAI-powered essay-writing assistant (EWA) can potentially guide students through key academic writing processes such as planning, drafting, and reviewing [7] without producing the full draft on their behalf. In this study, we envisioned EWA as a complementary learning assistant, particularly for independent study outside the classroom hours. Beyond simply supporting learners, EWA is designed to extend teacher support by providing scalable one-on-one writing guidance [27]. As a first step, this study aims to investigate student-GenAI interaction and their relationship with students writing performance, measured through essay scores.

## 1.1 Effective Tutor Conceptualization

As noted in the JISC report [15], there has been a shift regarding the use of GenAI from serving as a mere answer-provider to adopting the role of a collaborative learning assistant that emphasize dialogue in learning activities. One effective approach to facilitating such dialogue involves incorporating human-like interactive processes [5], particularly through the implementation of effective tutor behaviours. Applying such behaviours can enhance students' learning experience. One model that emphasize effective tutor behaviour is the INSPIRE Model, developed by Mark R. Lepper and Maria Woolverton [17]. This model identifies seven characteristics of effective tutors: Intelligent, Nurturant, Socratic, Progressive, Indirect, Reflective, and Encouraging. The Socratic component of the INSPIRE Model emphasizes the use of questions and hints rather than direct instructions and answers, enabling students to take the next step independently. This approach is widely recognized in literature as an effective tutoring strategy [6]. Although these strategies have primarily been studied in human tutoring contexts (both online and in-person), the INSPIRE Model itself is based on research by Lepper et al. [17] on elementary school tutors teaching mathematics, while Cukurova's study [6] examined one-on-one online tutoring. These findings present an opportunity to inform the development of adaptive GenAI tutoring systems that incorporate such strategies.

### 1.2 AI-Assisted Essay Writing

Natural language interaction with GenAI allows it to function as a feedback tool, providing tutoring support for students' writing tasks. This is similar to peer or tutor review in writing process approach [7], which emphasizes feedback provided by knowledgeable persons. Recent studies highlight GenAI's potential in supporting student writing. Imran et al. [12], in a systematic literature review of ChatGPT's role as writing assistant in higher education, synthesized that ChatGPT helped students in multiple aspects such as identifying grammar issues, improving writing structure, drafting, brainstorming ideas and summarizing literature. An empirical study by Punar et al. [21], investigating ChatGPT as a learning assistant for undergraduate students in Turkey, reported similar findings: ChatGPT supported self-editing by providing correction, making writing more formal, and offering explanations upon further questioning. Likewise, Susha et al. [25], through a case study integrating ChatGPT into the writing assignment in a Dutch University, examined how students used the tool and the challenges and benefits they experienced. Their thematic analysis of student reflections showed that undergraduate students benefited from ChatGPT's feedback in structuring essays, generating key arguments, researching a topic's main concepts and enhancing their academic writing processes.

However, GenAI can significantly influence both writing quality and students' thinking—opinionated AI, for instance, may shape how students construct their arguments in the essays [13]. These conditions often occur in proprietary GenAI tools such as ChatGPT, Claude, Gemini, Copilot, and DeepSeek, where the primary goal is productivity and efficiency. While these tools could assist students with writing, editing, formatting and summarizing documents, they prioritize task completion rather than pedagogical value [30]. As a result, they may limit learning opportunities for students to practice higher-order thinking and develop their writing skills, as the GenAI tools have already generated complete content [24, 25].

To address this, researchers have emphasized the importance of incorporating effective tutoring principles into AI systems. The INSPIRE Model [17] identifies essential features of effective tutoring practice, such as Socratic questioning and offering encouragement. Although these strategies emerged from human tutoring, research indicates that dialogue-based, adaptive AI systems could similarly motivate deeper engagement and foster student autonomy [15]. Building on these insights, the Essay Writing Assistant (EWA) used in this study is designed with pedagogical intent and contextual understanding to support students' essay writing. By integrating Socratic questioning [17] and effective tutoring behaviors [6], EWA aims to scaffold students through the writing process rather than simply generating drafts.

### 1.3 Student-GenAI Interaction in Writing Process

With the recent development of GenAI, student–AI interactions in the writing process have become increasingly central to discussions of learning and assessment. However, most prior studies have focused on what GenAI can accomplish for writing, whereas relatively few have examined how students interact with these systems across different phases of the writing process. Some empirical studies have started to investigate these interaction processes in greater depth.

For example, Nguyen, et al.[19] examined strategies employed by doctoral students when collaborating with ChatGPT. Using Hidden Markov Model (HMM) to analyze the sequence interactions and process mining to examine patterns, they found that students who engaged in iterative, highly interactive processes with the tool generally achieved better performance in the writing task. Similarly, Kim, et al. [16] investigated the interaction pattern of master's students with ChatGPT and their impact on academic writing by utilizing Epistemic Network Analysis (ENA) to capture differences between students with varying levels of AI literacy. Their results showed that students with high AI literacy collaborated more actively with GenAI in various writing phases (from planning to evaluating task outcomes). Moreover, these students achieved significantly higher scores across all rubric dimensions than their lower-literacy peers.

While these studies provide valuable insights into how students interact with general-purpose GenAI tools that prioritize task completion, less is known about how such interactions unfold when students work with pedagogically grounded systems, such as Socratic Model AI-Assisted essay writing tools. In particular, there is limited evidence on how different interaction patterns with pedagogically designed AI relate to students' writing quality. Moreover, prior research has primarily focused on doctoral and master's students, leaving undergraduate students underexplored.

This study addresses this gap by employing a mixed method design that combines quantitative and qualitative analyses to examine student interactions with GenAI and their relationship with writing quality. We use EWA, a GenAI powered essay writing assistant built upon pedagogical principles (INSPIRE model and effective tutoring behavior) to support undergraduate students in writing 500-1000 words argumentative essay during a summer school course. In the context of student interaction with EWA, this study aims to address the following questions:

- RQ 1: To what extent are specific patterns of student interaction during the writing process associated with variations in final writing quality across the four scoring dimensions (content, analysis, organization & structure, and quality of writing)?
- RQ 2: How do the unique student interaction patterns within high- and low- performing writing quality groups differ qualitatively?

## 2 METHOD

The objective of the study was to better understand student interaction with GenAI and their relationship with writing quality. EWA was employed as the primary tool to support students writing, and their work was evaluated using an argumentative essay scoring rubric. Bachelor's degree students from various majors participating in a university summer school were invited to take part in the experiment and completed 500-1000 words essay writing assignment in a two-hour session. The following method section details the study design, including the tools, participants and study context, research procedures, and data analysis methods.

### 2.1 Essay Writing Assistant (EWA)

We utilized an EWA prototype grounded on the academic writing process and effective tutoring behavior, leveraging system prompts to guide EWA in supporting essay writing effectively. These pedagogically grounded prompts serve as a guiding framework for the AI, shaping its behavior and ensuring that its outputs align with the intended goals [26]. Students can interact with the EWA prototype through three academic writing processes [9] and the characteristics of argumentative essays [7], including topic selection, outline development, drafting, and reviewing. The learning context of the writing tasks was also embedded in the EWA to provide contextual information and generate more targeted feedback for students. This included three core readings and the scoring rubric, as presented in Figure 1.

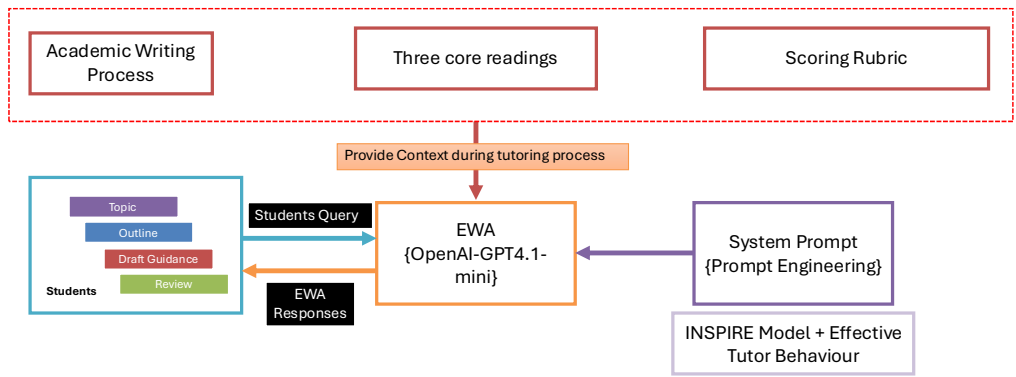

Figure 1. EWA Conceptual Model

The pedagogical prompt integrates the INSPIRE model [17] and effective tutoring principles [6]. Techniques include expert prompting to shape GenAI responses (e.g., *"You Are a Professor of AI in education and learning"*), and constraints to ensure adherence to instructions depending on the configuration conditions (e.g., *"Keep responses concise and focused, with a maximum of 150 words per reply"*). Few-shot, and chain-of-thought [2] prompting were used during the planning (topic selection and outlining) and content development (drafting) phase to guide reasoning, resulting in more coherent and accurate responses. Metacognitive prompting was applied in the review phase to emulate human introspection and regulation (e.g., *understanding the problem, identifying relevant concepts, formulating and evaluating a preliminary answer, and confirming the final response*). All techniques aimed to support critical thinking, preserve students' voices, and strengthen arguments without generating full drafts. We also embed additional information (core readings and scoring criteria) in the planning, content development (drafting), and review and feedback phases through the system prompt to ensure that GenAI responses are aligned within the contextual information. This supplemented the model's existing knowledge to produce more accurate responses. The system prompts used are available at **Appendices (1-2)**.

The web-based prototype, designed like the ChatGPT interface, was developed with an open-source Python framework—Streamlit (https://streamlit.io/), hosted on its cloud service as depicted in Figure 2. A welcome message introduced EWA's capabilities and role in supporting students essay writing, including an overview of its functions, the essay topic, and the core readings. On the left side, the interface displayed students chat histories, including the number of messages in each conversation. Students can flexibly interact with EWA by either following step by step guidance or focusing on specific tasks, such as rubric-based review. Google Firebase was used as the database to record these interaction logs. The GPT-4.1-mini API was selected as the backend GenAI service for its optimal balance between performance and cost-effectiveness. A detailed example of interactions log is available in **Appendix 3**.

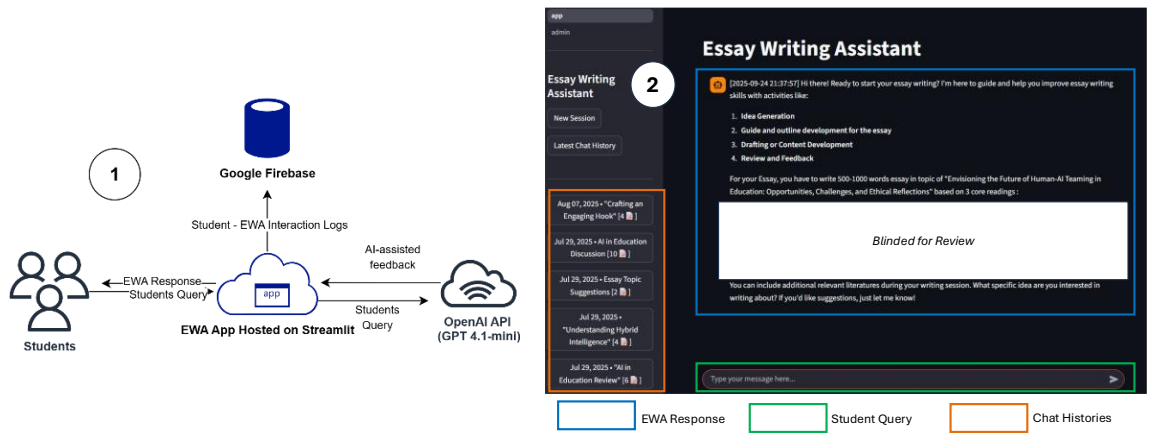

Figure 2. (1) EWA architecture, consisting of the EWA application hosted on Streamlit Community Cloud, interaction logs stored on Google Firebase, and the GPT-4.1-mini API as the GenAI backend. (2) EWA user interface displaying EWA responses, student queries, and chat histories.

## 2.2 Scoring Rubric

A rubric developed by Nguyen et al. [19] was adapted and modified to evaluate the quality of argumentative essays. In adapting and modifying the rubric, the goal was to ensure close alignment with the argumentative writing structure [4] and to allow flexible application across various topics. The rubric consists of five dimensions with scale from 1 (poor) to 4 (excellent): content (30%), analysis (30%), organization and structure (15%), quality of writing (15%), and word limit (10%), as illustrated in **Appendix 4**. 10% of the essays were anonymized and independently assessed by the three graders, resulting in Kendall's coefficient (W) = 0.89 which indicates a very high level of agreement among the three raters. The remaining essay was then randomly assigned to two graders to proceed.

## 2.3 Participants and Task Context

Fifty-six undergraduate students from various majors attending a university summer school in 2025 were invited to participate in the study. The institution approval along with students' consent was received prior to the study. The task was a part of the summer school assessment accounting for 40% of their final grade, where students engaged with 500-1000 words argumentative essay over two hours, on the topic *Envisioning the Future of Human-AI Teaming in Education: Opportunities, Challenges, and Ethical Reflections,* based on core readings. A relatively novel knowledge was chosen as the topic to control for prior knowledge across the cohort, while a two-hour lecture immediately prior to the exercise covered the key ideas of the topic. Students were allowed to include additional relevant references and interact with EWA at their disposal while no other AI assistants were allowed during their writing session. All participants were L2 English speakers who had already met the academic and English language entry requirements of the university summer school programmes, although they were non-native English speakers.

## 2.4 Research Procedure

Following the research procedure presented in Figure 3, the instructors introduced the tool and the content a day prior to the experimental day. The supported readings were assigned as additional readings for students. During this introduction session, students also reviewed an information sheet approved by the institutional board, indicated their consent and had access to EWA to familiarize themselves with the tool. On the following day, the instructors reviewed the content in a two-hour lecture with students to control prior knowledge and ensure topic familiarity. After that, a two-hours lunch break was provided to students to allow internalization of new knowledge, further preparation and to prevent fatigue and information overload immediately before the task. Students were then introduced to the task and the assessment criteria. They started engaging with EWA to prepare their writing assignments for two hours with no other GenAI tools, search engines and communication with peers allowed. Aside from EWA, students were only allowed to access the reading material and use X5Learn [20], an online open educational resource platform to search for information. All students were required to write an argumentative essay with the topic of "*Envisioning the Future of Human-AI Teaming in Education: Opportunities, Challenges, and Ethical Reflections*". The key data collected for this study were students' interactions log and their essay scores.

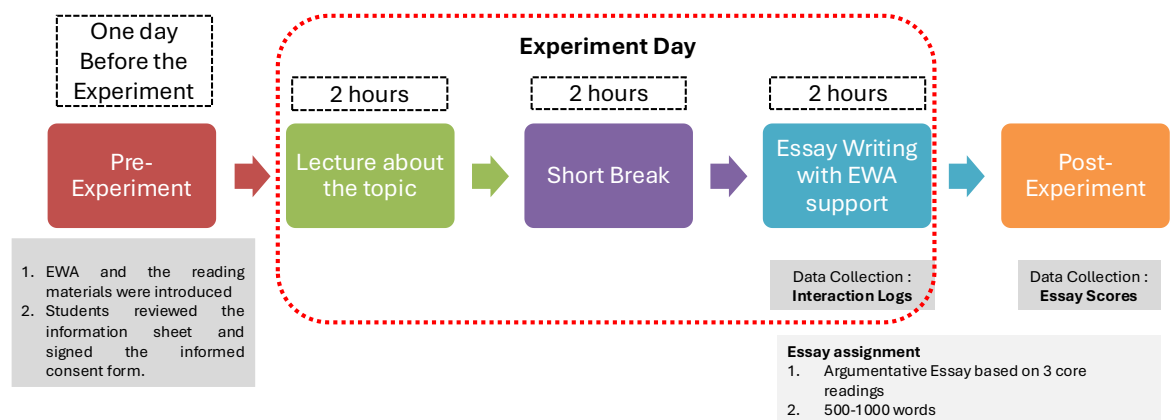

Figure 3. Research Procedure

## 2.5 Data Analysis

At the end of the study, 32 participants from an initial pool of 56 students completed the consent form, interacted with EWA, and submitted the essays. Each participant was assigned a pseudonym (e.g., 1, 2) to ensure anonymity. To further unpack how GenAI, as an essay writing assistant (EWA), can support students' learning in argumentative essay writing within the context of higher education, we analyzed their interaction pattern and final essay scores. To address the research questions, we examined both student interaction and writing performance through a combination of quantitative and qualitative analyses, using clustering, sequential pattern mining (SPM) and statistical analysis. For RQ1, we developed a codebook, applied SPM at the cohort level, and conducted statistical analysis of essay scores dimensions across clusters. For RQ2, we grouped essays by score, applied SPM within each cluster, identified unique patterns and further explained these unique patterns through manual qualitative content analysis. In this way, we aimed to capture a comprehensive picture of the relationship between interaction processes and students writing quality in terms of essay scores.

*2.5.1 SPM, Clustering and Statistical comparison of essay scores between clusters*

To address RQ1, we first developed a codebook to interpret student and EWA actions during essay writing. We followed the codebook development process outlined by Weston et al. [28], which included the phases of code conceptualization, application/review, and refinement. Student actions were categorized based on the academic writing process (deductive approach). An inductive layer was then added by identifying patterns that did not fit the predetermined codes and by creating new codes for emerging themes. The initial codebook development was conducted by the first and sixth researchers during a similar argumentative essay writing task in the 2024/2025 Education and Technology master's module. Furthermore, an Inter-Rater Reliability using Gwet AC1 was utilized to ensure that the researchers consistently applied the codebook criteria. Gwet AC1 was chosen to address the limitation of Cohen's Kappa in handling imbalanced codes and rater bias [29].

The first and third researchers conducted iterative refinement by applying codes to 10% of the labelled dataset, identifying gaps or overlaps, and revising both deductive and inductive codes until full agreement was achieved. Students' codes can be classified according to the writing process. In the planning phase, two codes were used: "*planning*" (p), given when students responded to EWA regarding their plan or next step, and "*asking help about planning*" (hep), indicating a help request on essay planning, including the outline. In the drafting phase, three codes were applied: "*writing*" (w), when students submitted part of their writing; "*adding references*" (m), when they inserted references or citations; and "*asking help about content*" (hec), when they requested help with essay content. In the review phase, three codes were also used: "*reviewing*" (r), when students revised their essay after interacting with EWA; "*asking help about review*" (her), when they sought help in reviewing their essay parts; and "*polishing*" (pl), when they asked EWA to polish their draft. In addition to these three phases, support-phase codes could appear in any stage of writing: "*agreement*" (a), when students expressed agreement with EWA's responses; "*disagreement*" (d), when they disagreed or argued against EWA's suggestions; and "*asking to translate*" (t), when they requested translation support (to or from English). The final code was "*ban*" (b), indicating students' intention to ask EWA to generate or write content for them. Altogether, this process resulted in twelve student codes, seven of which were derived deductively from the academic writing process [7] as presented in Table 1.

Inter-rater reliability was then calculated using Gwet AC1, and obtained an AC1 score of 0.97, indicating almost perfect agreement [29]. Any remaining disagreements were then discussed and resolved. The first researcher then labelled all essays in the study. The full labelling results, consisting of 1,282 labeled instances, are available in the **Supplementary Materials** and provide insight into how EWA is used in the context of argumentative essay writing.

Table 1. Student Codebook

| Action Codes | Definition | Codebook Development Types | Tutor/Student Behavior |
|---|---|---|---|
| p | Students respond to EWA with their next steps, including the topic, essay focus, outline and essay content. | Deductive (Theory Driven) | Planning |
| w | Students write essays, including an introduction, body paragraphs, and a conclusion. | | Drafting |
| m | Students add references or citations to their writing | | Drafting |
| r | Students revise their essays, including the introduction, body paragraphs, and conclusion. | | Revision |
| hep | Students seek EWA's help with essay planning | | Request for Feedback on Planning |
| hec | Students seek EWA's help with content-related questions. | | Request for Feedback on Drafting |
| her | Students seek EWA's help to review their work and provide feedback | | Request for Feedback on Reviewing |
| a | Students express agreement with EWA. | Inductive (Data Driven) | Planning/Drafting/Reviewing |
| d | Students argue with or express disagreement toward EWA. | | Planning/Drafting/Reviewing |
| t | Students seek EWA's help with translations (to and from English) | | Planning/Drafting/Reviewing |
| pl | Students ask EWA to polish their drafts | | Reviewing |
| b | Students ask EWA to write or generate content for them | | |

The next step involved identifying all student interactions using the codebook presented in Table 1. After that, SPM was conducted to identify unique behavioral patterns at cohort level (all participants). The PrefixSpan algorithm was utilized using Python to detect common interaction sequences based on the coded data [14]. The parameters were set as follows: maximum pattern length was left unrestricted, and the minimum support (minsup) was set 50%. This means a sequential pattern was considered frequent if it appears in more than half of all sequences. For instance, if a cluster contained 32 students, a 50% minsup required the pattern to appear in at least 16 students. The frequency of each frequent pattern was then counted on each student. K-means clustering was performed on these frequent patterns to classify students based on their interactions pattern. As the normality test indicated that all features had $p < 0.05$, a non-parametric test, specifically the Mann-Whitney U test was employed. This test was conducted to examine whether clusters of student interaction patterns differed significantly across the four essay scores dimensions that directly connected to the argumentative essay structure described in section 2.3 (content, analysis, structure & organization and quality of writing). SPSS version 29 was used during the statistical analysis.

*2.5.2 Unique Pattern Analysis*

To address RQ 2 on unique student interaction patterns, we employed qualitative analysis which involved grouping essays by score using the K-means clustering method, followed by SPM using PrefixSpan within each cluster. We then identified unique patterns in each cluster and further explained them through manual qualitative content analysis.

## 3 RESULTS

### 3.1 SPM, Clustering and Statistical comparison of essay scores between clusters (RQ1)

The interaction log of all 32 students in the same cohort was compiled, resulting in 1,282 labeled data points, with student interaction lengths ranging from 5 to 57 sequence. Applying the PrefixSpan method with a minsup values of 50% to the entire cohort resulted in 12 frequent patterns, as presented in Table 2.

Table 2. PrefixSpan Result of Interaction log in all students at 50% minsup.

| Pattern | Support |
|---:|:---:|
| hec | 29 |
| hec, hec | 24 |
| hep | 24 |
| hec, hec, hec | 22 |
| w | 21 |
| hep, hec | 20 |
| p | 19 |
| hec, hec, hec, hec | 18 |
| hec, hep | 17 |
| hep, hep | 17 |
| hec, w | 16 |
| hep, w | 16 |

**Legend:**

| | |
|---|---|
| hec | Asking help about content |
| hep | Asking help about planning (essay outline/structure) |
| w | Writing |
| p | Planning (Responding to the EWA about next step) |

With 12 patterns supported by at least 16 students, this threshold was considered sufficient to capture frequent patterns that could reveal differences across essay score clusters. Overall, students tended to interact with EWA more during the planning and drafting phases than during the review phase. This was indicated by the frequent occurrence of "*asking about planning*" (hep) and "*asking about content*" (hec), but none of "*asking to review*" (her). Moreover, (hec) was the most frequent sequence pattern, occurring as (hec), (hec, hec), (hec, hec, hec), and (hec, hec, hec, hec). This indicates that more than half of the students used EWA to ask content-related questions in a sequential manner.

All frequent patterns for each student were then counted with gaps allowed. This means that gaps between student actions were permitted when calculating how often each pattern occurred in a student's interaction session. The frequency of each frequent pattern per student is presented in **Appendix 5**. K-means clustering of the frequency of frequent patterns revealed two distinct clusters, as determined by the elbow method ($k = 2$). Table 3 presents the Means of each feature across the two clusters. In Cluster 1, students showed fewer "asking help about content" (hec) and showed fewer "*planning*" (p) initiatives but made more "*asking help about planning*" (hep). For instance, in the planning phase, some students asking about how to start and end the essay : "*so how can I write the start of the essay, my idea will be the Future of Human-AI Teaming in Education is promising and we will ultimately make good use of AI while there must be some obstacles*"(ID 21) and "*how to write the ending ?*" (ID 23). EWA responded with guiding questions, hints and short examples. Another type of question in planning phase concerned the essay outline, as asked by a student: "*How should I complete this paper?*"(ID 38). EWA responded with a step-by-step procedure starting from determining the focus, core argument, outline and writing steps. Similarly, a student asked an essay structure question: "*What could our structure look like, and does it need to involve opportunities, challenges, and ethical reflection?*" (ID 35). EWA responded by suggesting an introduction, body paragraph and conclusion, as shown in **The Supplementary Materials.**

While Cluster 2 consists of students who requested more "*asking help about content*" (hec) and showed more "*planning*" initiatives (p) but made slightly fewer "*asking help about planning*" (hep). In the drafting phase, students asked various questions categorized as "*asking help about content*" (hec), for instance, some students asked for content directly: "*What might be the ethical issues of AI in education?*" (ID 18) and "*why human-AI teaming can enhance teacher productivity via transactional and situational teaming*" (ID 6). One student asked for summarization: "*Summarize what the first article talked about*" (ID 28). Another student used EWA for brainstorming, asking*: "For the learner, too much AI using cause learners lose the ability to critically reflect, lose the spirit of innovation, and become rigid in thinking. What do you think of this idea?"* (ID 4).

Both clusters showed similar average active "*writing*" (w) frequency in the drafting phase. For the writing component (w), after students received guiding questions and hints about essay parts (introduction, body and conclusion), they submit their part of writing. For instance, after receiving guiding questions and hints about the introduction, a student was asked to write a draft of approximately 100 words and then submitted the following draft: "*In recent years, the integration of artificial intelligence and education has become a core issue in global educational reform. [............................]. This article will explore the opportunities and challenges of AI in education, emphasizing the irreplaceable role of teachers as "cognitive guides." This is the first part of my article. What do you think?*" (ID 35). Similarly, another student, after receiving hints from EWA about drafting a strong conclusion, wrote: "*this is my conclusion: In conclusion, the future of art education communication is undeniably intertwined with artificial intelligence. [..............]*" (ID 30). The full student-EWA interaction logs are available in the **Supplementary Materials**.

Table 3. Frequent Interaction Pattern Means per Cluster

| Frequent Pattern | Cluster 1 (Structure) | Cluster 2 (Content) |
|---|---|---|
| hec | 2 | 10 |
| hec, hec | 1 | 5 |
| hep | 3 | 2 |
| hec, hec, hec | 1 | 3 |
| w | 2 | 2 |
| hep, hec | 1 | 1 |
| p | 1 | 2 |
| hec, hec, hec, hec | 0 | 2 |
| hec, hep | 1 | 1 |
| hep, hep | 1 | 1 |
| hec, w | 1 | 1 |
| hep, w | 1 | 1 |
| N (each cluster) | 19 | 13 |

Legend:
- hec — Asking help about content
- hep — Asking help about planning (essay outline/structure)
- w — Writing
- p — Planning (Responding to the EWA about next step)

After applying Holm's correction for multiple comparison, no significant results ($\alpha = 0.05$) were observed in the four Mann Whitney U Test on student's essay scores (Content, Analysis, Organization, Quality of writing) across clusters (Table 4). Although no significant differences were found, in the Organization dimension Cluster 1 showed a higher Mean and Median score ($M = 3.42$, $Md = 4$, $N = 19$) compared to Cluster 2 ($M = 2.77$, $Md = 3$, $N = 13$), with a moderate effect size ($r = 0.36$).

Table 4. Mann Whitney U Test on Students' essay scores across clusters of interaction patterns

| Essay Score Dimensions | Mean_C1 | Mean_C2 | SD_C1 | SD_C2 | Median_C1 | Median_C2 | U | Z | p | p_adjusted (Holm) | r |
|---|---|---|---|---|---|---|---|---|---|---|---|
| Content | 3.11 | 3.31 | 0.66 | 0.63 | 3 | 3 | 103.5 | -0.86 | 0.39 | 1 | 0.15 |
| Analysis | 3.32 | 3.15 | 0.67 | 0.69 | 3 | 3 | 107.5 | -0.68 | 0.51 | 1 | 0.12 |
| Organization | 3.42 | 2.77 | 0.69 | 0.73 | 4 | 3 | 74.5 | -2.03 | 0.04 | 0.16 | 0.36 |
| Quality of Writing | 3.53 | 3.54 | 0.51 | 0.66 | 4 | 4 | 120.5 | -0.13 | 0.91 | 1 | 0.02 |

## 3.2 Unique Pattern Analysis (RQ2)

K-means clustering of the essay scores dimensions (Content, Analysis, Organization, and Quality of Writing) revealed two distinct clusters, as determined by the elbow method (k = 2). Table 5 presents the Mean of each essay score dimension across the two clusters.

Table 5. Essay Score Means per Cluster

| Clusters | Content | Analysis | Organization | Quality of Writing | N | Group Description |
|---|---|---|---|---|---|---|
| 1 | 3.4 | 3.4 | 3.6 | 3.7 | 21 | Higher Writing Quality |
| 2 | 2.7 | 2.8 | 2.5 | 3.1 | 11 | Lower Writing Quality |

PrefixSpan was applied to each cluster using minsup values ranging from 50% to 70%, followed by the identification of unique sequences. Unique patterns were found at minsup 65% with three patterns shared by both groups, one unique pattern in Cluster 1 and two unique pattern in Cluster 2. The results are presented in Table 6.

Table 6. PrefixSpan results for higher and lower writing quality groups at 65% minsup

| Pattern | Cluster 1 (Higher Writing Quality) | Cluster 2 (Lower Writing Quality) |
|---|---|---|
| hec | 19 | 10 |
| hec, hec | 15 | 9 |
| hep | 15 | 9 |
| w* | 14 | 0 |
| hec,hec,hec** | 0 | 9 |
| hep,hec** | 0 | 8 |

Legend:
hec    Asking help about content
hep    Asking help about planning (essay outline/structure)
w      Writing

\* one unique pattern in higher performance group
\*\* two unique patterns in lower performance group

The result indicated that "*asking help about content*" (hec), (hec,hec) and "*asking help about planning*" (hep) were three patterns that frequently occurred in both clusters. Higher writing quality was indicated by using EWA more actively to enhance their essay quality. Unique patterns in the higher writing quality group included student action of "*writing*" (w). This pattern suggests that students consistently submitted portions of their writing, such as the introduction, body paragraphs, or conclusion, to EWA for feedback (w). After responding to EWA's queries and requesting clarification when needed (hep and hec, which occurred in both clusters), students in higher writing quality groups actively wrote and shared sections of their essay with EWA for further guidance. None of these active writing patterns appeared among the most frequent sequences in the lower writing quality group. In contrast, the lower writing quality group exhibited frequent subsequences such as (hec,hec,hec) and (hep,hec). These patterns suggest that students interacted with EWA in a more passive manner, often "*asking help about planning*" (hep) or asking help about content (hec), without taking further initiative. After responding to EWA's queries, they tended to ask additional questions but did not follow up with active writing, such as drafting or submitting parts of their essays for feedback.

Manual Qualitative content analysis confirmed the identified patterns in each cluster. For the higher writing quality group, two interaction logs were carefully analyzed: Student IDs 20 and 21. Student ID 20 began with "*planning*" (p) action in respond to EWA's questions. The student then spent several rounds of interaction shaping their outline by "*asking help about planning*" (hep) and "*asking help about content*" (hec). Subsequently, the students submitted part of their essay, including the introduction, body paragraphs and conclusion (w), and gradually refined them based on EWA's feedback. Similarly, Student ID 21 began by "*asking help about content*" (hec), specifically requesting a summary of the three core readings. The student later "*asking help about planning*" (hep) by inquiring how to draft the introduction and body paragraphs, including brainstorming examples and counterarguments. Finally, the student submitted writing paragraph by paragraph, including the introduction, body paragraphs and conclusion (w) and "*ask to polish*" (pl) the initial draft. Excerpts from the action pattern of Student IDs 20 and 21 are provided in Table 7.

Table 7. Student action patterns in the higher writing quality group

| Student ID 20 | | Student ID 21 | |
|---|---|---|---|
| Label | Interaction Log | Label | Interaction Log |
| p | "I am interested in teachers' initiative in STEM education in the AI era. Please help me organize my writing outline." <br> EWA: responded with guiding questions and hints for structuring the introduction, body paragraph and conclusion. | hec | "so what the core idea of these 3 papers" <br> EWA: summarized three core readings |
| p, hep | "My proposed framework for this introduction: [...............] Please review my framework and see if there are any additions, refinements, or deletions that need to be made." | hep | "So how can i write the start of the essay, my idea will be the Future of Human-AI Teaming in Education is promising and we will ultimately make good use of AI while there must be some obstacles." <br> EWA: responded with guiding questions and hints for structuring the introduction and provided a short example |
| | [.......................................................] | hep | "How about the body of the essay" <br> EWA: responded with guiding questions and examples for structuring the body paragraphs |
| w,her,pl | "Paragraph 1: With the rise of artificial intelligence, the field of education is undergoing profound changes. ...... Paragraph 2: First, teacher agency is indispensable in maintaining professional judgment. For example, [......] Paragraph 3 (Further Reflection): Furthermore, I believe that the development of AI has actually fostered teacher agency in STEM education. [..................] Paragraph 4 (Conclusion): In summary, [........] Please help me revise and polish the third and fourth paragraphs." | hec | "so, for each point did you have some examples" <br> EWA: responded with examples from the core readings. |
| | [.......................................................] | | [.......................................................] |
| w, pl | "I need to add that before the fourth paragraph above, I should write another paragraph on how to use AI correctly: [.................] Please help me improve this description." | w,pl | "With the development of artificial intelligence, AI has completely entered our lives. [...........] polish it" |
| | | w,pl | "First and foremost, there is no denying that AI is more clever than human in most instances. [...........] (Lawrence et al., 2024). polish it" |
| | | w,pl | "On the one hand, erudite AI may teach better than normal teachers to some degrees. [.........] polish it" |
| | | w, pl | "However, challenges and risks that come with AI are inevitable. In my university, [................] polish it" |

For the lower writing quality group, we closely examined Student IDs 24 and 31. Both students showed a similar pattern of repeatedly "*asking help about content*" (hec). Although they appeared engaged with the essay and the core reading content, they did not follow up their questions by submitting written sections for feedback. They continued to seek help with planning and content iteratively but did not submit any actual writing for feedback. Excerpt from the action patterns of students in the lower writing quality group are presented in Table 8.

Table 8. Student action pattern in the lower writing quality group

| Student ID 24 | | Student ID 31 | |
| --- | --- | --- | --- |
| Label | Interaction Log | Label | Interaction Log |
| hec | "Please explain the definition of human-machine collaboration and the advantages of human-machine collaboration" | hec | "Hello, EWA, could you tell me your understanding of the ethical reflections?" |
| hec | "What difficulties in education can be aligned with the advantages of human-machine collaboration?" | hec | "Please tell me the typical theory of AIED's future development said by deans." |
| hec | "What exactly do we mean by cognitive overload, limited human perception, or scalability?" | | [..........................................................................] |
| | [..........................................................................] | p, hep | "If I write an essay of the coexistence of challenges and advantanges, the theme maybe not good?" |
| hec | "what's the Future of Human-AI Teaming in Education" | hep | "ok, please tell me the main idea of the essay" |
| | | hec | "What is the confliction between teachers and AI" |
| | | hec | "How about students and AI" |

## 4 DISCUSSION

### 4.1 Student Interaction Pattern and Essay Score Comparison

The frequent interaction patterns at the cohort level were most observed during the planning and drafting phase. These patterns reflected a combination of active and passive behavior. Passive behavior was characterized by asking questions without contributing ideas, following up or submitting written work. In contrast, active behavior was characterized by students' contributions to essay development. For passive behavior, we observed the instances of "*asking about planning*" (hep) and several sequences of "*asking about content*" (hec) within the identified frequent pattern. Such tendencies may be attributed to the EWA design, which specifically instructed to support students in essay planning (including outline and structure), drafting (incorporating three core readings to provide contextual answers), and reviewing (providing feedback based on rubric criteria). This behavior aligns with previous research demonstrating the benefit of GenAI in supporting students essay writing during the planning phase [16], particularly in developing their essay outlines and structures [25].

The findings can be analyzed from two perspectives. First, student asked content-related questions to help them understand certain topics, echoing the findings of Punar, et al [21] in their empirical study about the role of ChatGPT as learning assistant. In their research, students mentioned "*clarifying doubts and providing explanations upon further questioning*" as one of the perceived benefits of ChatGPT as a learning assistant. Students also used EWA to brainstorm and summarize, indicating that EWA can support various aspects of the writing process, as reported by Imran, et al [12] in their systematic literature review on ChatGPT's role as writing assistant. One of the themes in their findings was that ChatGPT can provide significant assistance in generating text, drafting initial versions, brainstorming ideas and summarizing literature. Second, in all content-related questions, EWA responded based on the specific contextual information provided in the system prompt (core readings), which can be verified in the **Supplementary Materials**. As a result, EWA's responses were more contextual compared with those of general-purpose GenAI. These responses indicated EWA's effort to address common limitations of standard GenAI such as inaccuracies in AI-generated feedback [1] and misalignment with course-specific standards [10]. At the end of each response, EWA provided instructions or questions (depending on the writing phase) that prompted students to reflect on the feedback and consider whether it aligned with their own thinking and analysis. Regarding active student behaviors, we found only "*planning*" (p) and "*writing*" (w) in the most frequent pattern, however, this does not mean that there are no active sequences of "*adding* *references*" (m) and "*revision*" (r). Instead, PrefixSpan did not detect these patterns under the 50% minsup threshold. Both the planning and writing actions demonstrated active engagement with EWA.

The effect sizes of Organization dimension ($r$=0.36) in Mann-Whitney test yielded a more nuanced interpretation of the findings. Both groups achieved mean scores above 3 on the 4-point rubric (see Table 4) for Content, Analysis, and Quality of Writing, suggesting that students consistently produced high quality essays in these areas regardless of how they interacted with EWA. However, the trend observed in Organization highlights that structural aspects of writing may be more sensitive to students' engagement with EWA. Students in Cluster 1 scored slightly higher in Organization, which may indicate that their interaction strategies contributed to more coherent essay structures, although the difference did not remain statistically robust after correction. Several factors may account for the absence of a significant difference. First, the relatively small sample size (19 vs 13 students per cluster) and relatively short interaction durations (2 hours) limited statistical power. Second, the variability within groups combined with the 4-point rubric scales may have reduced the ability to detect subtle differences. Finally, it is possible that EWA support equalized performance across clusters, providing sufficient support for all students to reach a similar standard of essay quality.

### 4.2 Unique Pattern of Higher and Lower Writing Quality Groups

Differences in students' frequent interaction pattern emerged between higher and lower writing quality groups. Our findings revealed that students in the higher group interacted in a more active and iterative manner to enhance their essay quality. After responding to EWA's queries and requesting explanations when needed, the higher group actively wrote and shared their essay sections with EWA for further guidance. In contrast, the lower group interacted with EWA in a more passive manner, often seeking help with planning or content without taking immediate active writing initiatives. These results align with the findings of Sabourin et.al. [22] on behavioral patterns of self-regulated learners in virtual environments, which concluded that learners exhibit different sequences of behavior depending on their degree of self-regulation. It could be that students in the higher group also possessed stronger self-regulation skills. They tended to gather information and immediately record it carefully, whereas the students in the lower group tended to exhibit more passive behavior. Furthermore, active engagement has previously been linked to the knowledge construction process, which contributes to improvements in writing quality, as empirically demonstrated by Nguyen [19] and Yang [31] in their studies on GenAI-assisted writing. In Nguyen's study, higher performing doctoral students demonstrated strategic and dynamic interactions with GenAI. Similarly, Yang's findings highlighted that writers who frequently modified GenAI-generated text, thereby engaging in higher-order cognitive processes, consistently improved essay quality. Since we only analysed students' interaction logs with EWA, there could be other factors that could influence their interactions which were not measured in this study. For instance, this could reflect delayed behaviors of students in lower writing quality group, who may not have engaged with feedback promptly, resulting in a lack of writing content to be evaluated with EWA. In this case, our findings warrant further investigation, such as incorporating post-experiment interviews to analyze and clarify students' interaction behavior.

## 5 CONCLUSION AND STUDY IMPLICATION

This study makes several contributions. Methodologically, this study builds upon and extends the INSPIRE model and the principles of effective tutor behavior from the human tutoring context (in person or online) to steer GenAI through pedagogical prompting in the essay writing context. The EWA system prompt can be flexibly adapted to other educational contexts. We also developed a codebook to label student interactions based on the writing process approach. Drawing on this codebook, we analyzed student interaction patterns and their corresponding essay scores using both quantitative and qualitative approach. Empirically, findings from quantitative approach highlighted that students primarily focused on planning or content development, both of which can be effectively supported by EWA through contextual responses. Moreover, the moderate effect size in the Organization dimension indicates that students' interaction strategies may have contributed to more coherent essay structures, although the difference did not remain statistically significant after correction. The qualitative approach suggested that students in the higher writing quality group were more likely to submit their essays to receive feedback from EWA. Our findings suggest that teachers could encourage active engagement patterns, such as submitting full essays or partial drafts after asking EWA questions about planning or content. This approach allows students to receive feedback from EWA based on a predetermined scoring rubric, which they can then choose to accept, modify, or reject. Future EWA development could integrate automatic labeling and monitoring of interactions to prompt students to move from asking questions to writing essay sections. Such practices may strengthen their essay writing processes and enable them to benefit more fully from GenAI-supported learning environment.

## 6 LIMITATION AND FUTURE WORK

However, this study has several limitations that future research should address. First, the dataset was limited to a single session with 32 undergraduate students performing argumentative writing tasks, which may not capture broader writing behaviours. Additionally, the four-point rubric may have constrained sensitivity to subtle differences. Future research could build on these findings through multi-session studies, varied learning tasks, larger samples, and more fine-grained measures. Furthermore, this study focused only on student interaction patterns and did not consider the semantic similarity of students writing. Further research should adopt a more holistic examination of the learning process, including both interaction logs and semantic similarity analyses between students writing and EWA responses, to validate and extend these initial findings and to provide deeper insights into how students engage with EWA. Another limitation is that in the current study, EWA was used solely as support tool to enhance essay writing based on predetermined rubric scoring and a Socratic approach. Subsequent studies could investigate using EWA as a training or learning tool, evaluated through pre- and post- test assessment in which students write essays without assistance from the tool. Overall, addressing these limitations offers opportunities to refine GenAI-powered writing assistants such as EWA for broader educational contexts. Such efforts can help ensure that they effectively address concerns such as over-reliance, which may lead to cognitive atrophy and reduced creativity, while at the same time optimizing the benefits of GenAI to complement teacher support and augment students learning.

## A. Appendices
[LAK 26-Appendices](LAK 26-Appendices)

## B. Supplementary Materials
[Student EWA Interaction Log](Student EWA Interaction Log)